\documentclass[12pt,a4paper]{article}

\usepackage[british]{babel}

\usepackage[a4paper,top=2cm,bottom=2cm,left=2.5cm,right=2.5cm,marginparwidth=1.75cm]{geometry}

\usepackage{amsmath, graphicx, hyperref, booktabs, caption, placeins, rotating, dcolumn,authblk, setspace, titlesec, lineno, listings, multirow, float, array, siunitx,tablefootnote}

\usepackage{natbib}
\title{Survey Experience and Nonresponse in an Online Probability Panel: A Survival Analysis\thanks{Acknowledgement: We are grateful to Joel Williams and Verian Group UK Ltd for providing the data used in this article and answewring our many questions about it.}}

\author[1]{Katya Kostadintcheva, Jouni Kuha, Patrick Sturgis}
\affil[1]{Department of Methodology, London School of Economics}
\date{}
\begin{document}
\maketitle

\begin{abstract}
We fit discrete-time survival models to data from an online probability panel (OPP), where the outcome is the respondent's first nonresponse to a survey invitation, following at least one previous survey completion. This approach has the advantage of utilising information about survey experiences over multiple survey waves, while accommodating the unbalanced data structure typical of OPPs, where the number, timing and content of survey invitations varies widely between panel members. We show that the nature and quality of previous survey experience has a strong influence on the propensity to respond to the next survey invitation. Longer surveys, reporting a survey as less enjoyable, a phone interview, and more days since the last survey invitation are found to be important predictors of nonresponse. We also find strong effects of personality on response propensity across survey invitations. Our findings have important implications for survey designers wishing to minimise nonresponse and attrition from OPPs. .
\end{abstract}

\section{Introduction}
Online probability-based panels (OPPs) play an increasingly important role in the modern survey research and opinion polling landscape. As with online surveys in general, OPPs gained prominence at a time of increasing costs and declining response rates to interviewer-administered surveys \citep{callegaroOnlinePanelResearch2014}. At the same time, near-universal internet coverage in many countries, allied with rapid technological developments in online digital devices, combined to make web surveying the dominant survey mode \citep{mcpheeDataQualityMetrics2022}. The typical design of an OPP involves drawing a random sample from population or address registers and then contacting the sample, using offline methods, with an invitation to complete an initial survey online. This 'recruitment survey' is used to invite respondents to join the panel and to collect basic demographic and attitudinal data about them \citep{blomComparisonFourProbabilityBased2016, boschUtilityProbabilitybasedOnline2023}. \footnote{Another option is to recruit the panel off the back of an existing interviewer-administered survey. The first OPP in the UK, the NatCen panel, for example, was recruited using the face-to-face British Social Attitudes survey, with respondents invited to join the panel at the end of the interview \citep{jessopDevelopingNatCenPanel2017} The very high costs of interviewer-administered surveys is making this approach increasingly rare however.} Once panel members are recruited, the large majority complete subsequent surveys online, with  alternative modes, usually paper questionnaires or telephone interviews, for those who lack, or do not want to use, the internet (referred to as 'offliners'). The infrastructure for implementing and maintaining OPPs is now well-established in many countries in Europe, North America, Asia and Australia \citep{kaczmirekBuildingProbabilitybasedOnline2019}.

The appeal of OPPs is obvious. Using random sampling designs, online self-completion and alternative modes for offliners, they offer a high-quality, fast, and cost-effective vehicle for surveying general populations and sub-groups. Although they cost significantly more than nonprobability panels, OPPs yield markedly higher data quality due to their use of probability sampling methods and their ability to prevent bots and fake respondents from completing surveys \citep{kennedyAssessingRisksOnline2020,lugtigNonresponseAttritionProbabilitybased2014a, mercerComparingTwoTypes2023}. 

A significant challenge faced by OPPs, however, is the very low response rates they attain, particularly compared to face-to-face interview surveys. Push-to-web surveys are the most common method of recruitment to OPPs and these are themselves characterised by low response rates, generally in the range 10-35\% in the UK, depending on the design \citep{jessopComparingFacetofaceOnline2021,kantarpublicCommunityLifeSurvey2021,williamsHowWeBuilt2022}. Some respondents to the recruitment survey then decline the invitation to join the panel, while some of those who do agree to join fail to respond to individual survey invitations, or attrit from the panel entirely \citep{mcpheeDataQualityMetrics2022}. Combining these different sources of nonresponse produces net response rates of around 5 percent or less \citep{jessopComparingFacetofaceOnline2021}. Response rates this low raise legitimate concerns about the possibility of large biases in survey estimates, even after nonresponse and calibration weighting. Understanding the factors that predispose panel members to decline survey invitations in OPPs, as well as how these can be mitigated, is therefore a key challenge for survey practitioners. 

The number of active panel members in OPPs tends to be substantially larger than the required sample size for most individual surveys. For this reason, the majority of OPP surveys invite a random subset of panelists. This means that, over time, different panelists will have received survey invitations at different intervals and completed different numbers of surveys. The sets of surveys that they have completed will also vary widely in terms of topic coverage, survey duration, the types of questions administered, and so on. This feature of OPPs yields an analytical advantage for the study of nonresponse because it induces variability in the previous survey experiences of those invited to take part in any one survey. In this paper, we exploit this property, using paradata and self-reports to assess how different aspects of survey experience affect the probability of nonresponse to the next survey invitation. To foreshadow our later results, we find that self-reported survey enjoyment, the number of days since the last invitation, survey mode, survey duration, and the length of time each respondent takes to complete the survey influence subsequent survey response decisions. We also find that personality is important, with panelists' scores on some dimensions of the Big Five personality inventory emerging as strong predictors of nonresponse across survey invitations. 

The remainder of the paper is structured as follows. The next section reviews existing evidence on the impact of survey experience, burden, and personality in determining response decisions. We then describe the data and analytical approach before presenting the empirical results. The final section discusses the implications of our findings for survey practice, considers the limitations of our approach, and suggests fruitful avenues for future research.

\section{Relevant Literature}
While OPPs are most commonly used for implementing cross-sectional surveys, they differ in important respects from one-off push-to-web surveys, in that most respondents will already have completed one or more surveys when they are invited to complete a new one \citep{mcpheeDataQualityMetrics2022}. Thus, previous survey experience becomes potentially important in understanding responses to subsequent invitations. Research on the downstream effects of survey experience has drawn on the concept of 'burden' as an umbrella term for a range of different dimensions of how respondents experience the completion of surveys \citep{kaplanPrefaceOverviewSpecial2022}. The key point from this literature, perhaps an obvious one, is that the more burdensome a respondent finds the survey experience, the more likely they will be to yield poor quality data, break off before completing the questionnaire, and decline to respond to future requests for participation.  

Norman Bradburn provided the first exposition of the concept, describing burden as “the product of an interaction between the nature of the [survey] task and the way in which it is perceived by the respondent” \citep[p.~36]{bradburnRespondentBurden1978}. Burden is commonly measured using both objective and subjective indicators \citep{kaplanPrefaceOverviewSpecial2022}. Objective measures are derived from paradata on survey design features, such as response times, question complexity indicators, and item nonresponse counts \citep{rossmannUsingParadataPredict2015}. These 'passive' indicators have the benefit of being easy to collect and measure as they do not require respondent input. Subjective measures of burden, on the other hand, rely on respondent self-reports of survey experience, capturing “negative feelings or hardships experienced by survey respondents” \citep[p.~940]{yanResponseBurdenReview2022}. Common subjective indicators of burden ask respondents to assess whether they enjoyed doing the survey, whether they thought it was interesting, too long, too sensitive, and so on.  

There is a large literature assessing how survey experience and burden are related to nonresponse \citep{yanResponseBurdenReview2022}. Though differing in their specifics, these studies typically regress measures of panel nonresponse at time $t$ on survey experience at $t - 1$, controlling for respondent characteristics. Such studies often employ pooled and wave-on-wave regressions or regressions with aggregated panel response outcomes. While they generally use more than one wave of data, they are cross-sectional in nature as they use only between-person variability in the variables of interest. Scholars adopting this strategy have found that high participation rates and a mix of phone and self-completions in previous waves is associated with a higher probability of response at later waves \citep{kocarPowerOnlinePanel2023,lugtigUsingParadataExplain2018, tortoraAttritionConsumerPanels2009}. \citet{lugtigUsingParadataExplain2018} also found panel attrition to be lower among panelists who reported that they liked the previous survey, found it interesting and did not find the questions personal. Counter-intuitively, nonresponse in this study was lower for panel members who left a negative comment about the survey \citep{lugtigUsingParadataExplain2018}. 

In contrast, higher rates of item nonresponse, a larger gap between previous survey invitation and survey completion, and longer panel membership have been found to be associated with nonresponse and attrition \citep{lugtigUsingParadataExplain2018, rossmannUsingParadataPredict2015,tortoraAttritionConsumerPanels2009}. No association was found between attrition and previous smartphone use, reminders, or multiple completion sessions recorded at the previous wave \citep{lugtigUsingParadataExplain2018}. Using a latent class model to summarise differences in attrition patterns across multiple OPP surveys, \citet{struminskayaDataQualityProbabilitybased2014} found that panel members who perceived a survey negatively were less likely to belong to the class of committed panel members. The findings on the effect of previous survey duration are mixed, with \citet{lugtigUsingParadataExplain2018} reporting no impact and \citet{rossmannUsingParadataPredict2015} finding a negative effect on subsequent response. Using an experimental design, \citet{lynnLongerInterviewsMay2014} found no effect of survey length on subsequent response propensity in a face-to-face panel, although the time between survey waves was considerably longer in this context than is the norm for OPPs. 

A smaller set of studies have exploited within-person variability from repeated measures to understand time-varying survey experience factors, using methods of panel data modeling such as fixed effects, random effects or Latent Markov models. \citet{jinRelationshipSurveyBurden2022} found, for the Understanding America Study OPP, that individual questionnaire length (measured by number of screens viewed) was a positive predictor of subsequent nonresponse. They found no relationship between the number of surveys taken and subsequent response propensity. \citet{kocarPowerOnlinePanel2023} analysed data from the Life in Australia OPP, finding that nonresponse was lower among panel members who had completed more surveys and who had lower non-contact and refusal rates. Using the GESIS OPP, \citet{minderopPredictingPanelAttrition2023} found that longer gaps between survey invitation and completion were associated with higher attrition. \citet{gummerNoteHowPrior2020} also used the GESIS panel, to estimate the effect of a set of subjective respondent evaluations of the completed survey on subsequent response. They found that participation was lower when prior surveys were perceived as too personal (online mode only), too difficult (mail mode only) or too long (both modes). Likewise, \citet{struminskayaDataQualityProbabilitybased2014} examined attrition through a range of panel data models, finding that nonresponse was lower when the previous survey was rated as important for science and higher for burdensome previous experience. Incentive levels did not influence panel attrition after controlling for redeeming the incentive. \citet{roscheSurveyAttitudeIndicator2020} also found general survey attitudes (rather than perceptions of specific surveys) to be predictive of nonresponse using the LISS OPP, with panelists who rated surveys as enjoyable, valuable for society, and as less burdensome being more likely to respond. 

While panel data models with fixed or random effects are useful for examining relationships between survey experience and nonresponse using within-person variability, they typically do this without considering the timing of nonresponse, or how the role of the predictors changes over time. Survival analysis is better suited for these kinds of questions. Only two studies have adopted this approach to date. \citet{schwerdtfegerHowDoesBroadband2024} applied Cox proportional hazard models to the GESIS OPP to assess how spatial heterogeneity in broadband access is related to mode choice and panel attrition. No significant link was found between broadband supply and panel attrition but panel dropout was higher when respondents rated the previous surveys as long and when the measured duration was long. However, these variables were time-invariant in this study, so did not capture variability in survey experiences over time. \citet{roscheSurveyAttitudeIndicator2020} also used survival analysis, implementing a discrete-time survival model on the LISS panel. They found that the Survey Attitudes Scale was a stronger predictor of dropout than socio-demographic variables; the more enjoyable, socially valuable and less burdensome surveys in general were perceived to be, the lower the hazard of attrition. As with the random and fixed effects models in their study, however, this tells us little about how experiences and perceptions of specific surveys affect subsequent response decisions.

Our objective in this study is to build on and extend this body of evidence on how burden and other features of the survey experience are related to subsequent nonresponse. We fit discrete-time survival models to a mixed-mode OPP, where the outcome is first nonresponse, using a range of time-varying survey experience predictors, including objective and subjective measures of  burden and controlling for survey design and individual differences. 

In addition to burden and survey experience, we also examine how the likelihood of nonresponse is related to a respondent's personality, specifically the `Big Five' personality traits of 
Extraversion, Openness, Neuroticism, Conscientiousness, and Agreeableness  \citep{mccraeFiveFactorTheoryPersonality1999}. A growing body of evidence suggests that personality is an important factor in determining survey nonresponse and attrition \citep{sassenrothImpactPersonalityParticipation2013}. It can influence how different respondents evaluate the same set of experiences with a survey and thus affect how those experiences translate into nonresponse decisions. However, although personality seems important, the existing literature presents mixed findings regarding how different personality dimensions influence nonresponse propensity. This is partly due to the theoretical expectation being unclear for some of the dimensions. The expectations are clearest for Conscientiousness and Agreeableness, which are expected to be correlated with lower nonresponse due to a greater sense of responsibility, commitment and reliability for Conscientious people and a stronger sense of civic duty, cooperation and willingness to engage in social situations for those who score higher on Agreeableness. 

In terms of empirical findings, higher Conscientiousness is, as expected, generally linked to lower nonresponse and attrition \citep{chengPersonalityPredictorUnit2020a, lugtigNonresponseAttritionProbabilitybased2014a,satherleyDemographicPsychologicalPredictors2015a}, although some studies have found no association \citep{hanssonCanPersonalityPredict2018a,richterPersonalityHasMinor2014,salthouseSelectivityAttritionLongitudinal2014}. Higher Extraversion has been found to correlate with panel attrition \citep{hanssonCanPersonalityPredict2018a,lugtigNonresponseAttritionProbabilitybased2014a,salthouseSelectivityAttritionLongitudinal2014, satherleyDemographicPsychologicalPredictors2015a} but no association between this dimension and nonresponse was found by \citet{chengPersonalityPredictorUnit2020a} and \citet{richterPersonalityHasMinor2014}. Results for Agreeableness and Openness are more mixed. Some studies have found a link between high Agreeableness and greater risk of panel attrition \citep{lugtigNonresponseAttritionProbabilitybased2014a, roscheSurveyAttitudeIndicator2020}, while others have found effects in the opposite direction \citep{hanssonCanPersonalityPredict2018a,richterPersonalityHasMinor2014,salthouseSelectivityAttritionLongitudinal2014} and some report no significant effect \citep{chengPersonalityPredictorUnit2020a, satherleyDemographicPsychologicalPredictors2015a}. Openness has also been linked to both lower cooperation \citep{chengPersonalityPredictorUnit2020a,lugtigNonresponseAttritionProbabilitybased2014a,satherleyDemographicPsychologicalPredictors2015a} and higher cooperation \citep{richterPersonalityHasMinor2014, salthouseSelectivityAttritionLongitudinal2014}, and some studies finding no significant association \citep{hanssonCanPersonalityPredict2018a}. Negative emotionality (Neuroticism) has mostly been found to be unrelated to survey participation \citep{chengPersonalityPredictorUnit2020a,lugtigNonresponseAttritionProbabilitybased2014a,richterPersonalityHasMinor2014,salthouseSelectivityAttritionLongitudinal2014} although some studies have found it to be weakly predictive of attrition \citep{hanssonCanPersonalityPredict2018a, satherleyDemographicPsychologicalPredictors2015a}. 

\section{Data: Verian Public Voice online probability panel}
We use data from the Verian Public Voice (PV) OPP in the UK (see \cite{williamsKantarPublicVoice2022}, for details of the methodology and processes of the PV survey). Its target population is UK residents aged 16+, living in private households. The PV panel is periodically refreshed using new rounds of recruitment. The data that we use here is for respondents recruited from three separate recruitment surveys carried out between 2019 and 2021 (Table \ref{tab:recruitment}). The first of them was conducted using a combination of a face-to-face and a larger push-to-web survey. The subsequent recruitment surveys in 2020 and 2021 used the push-to-web protocol only. The proportion of the push-to-web surveys that were nevertheless completed by paper questionnaire was 20\%, 15\% and 21\% in recuitment cohorts 1, 2, and 3 respectively. All three recruitment surveys included a conditional incentive of a £10 gift voucher for participation. The response rates for each recruitment survey and the proportion who agreed to join the panel are shown in Table \ref{tab:recruitment}.

\begin{table}[h]
    \centering
    \caption{Recruitment survey dates, response rates, and panel registration rates}
    \label{tab:recruitment}
    \renewcommand{\arraystretch}{1.2} 
    \begin{tabular}{clccccc}
        \hline
        \multirow{2}{*}{\textbf{Cohort}} & \multirow{2}{*}{\textbf{Method}} & \multirow{2}{*}{\textbf{Fieldwork Dates}} & \multicolumn{2}{c}{\textbf{Response Rate}} & \multicolumn{2}{c}{\textbf{Panel Registration}} \\
        & & & $\mathbf{N}$ & \textbf{\%} & \textbf{N} & \textbf{\%} \\
        \hline
        \multirow{2}{*}{1} & Face-to-face & Sep 2019 – Jan 2020 & 1,630 & 35.8 & 1,323 & 81.2 \\
        & Push-to-web & Sep – Nov 2019 & 2,306 & 7.2 & 2,006 & 87.0 \\
        \hline
        2 & Push-to-web & Apr – Sep 2020 & 9,447 & 9.3 & 8,075 & 85.5 \\
        3 & Push-to-web & May – Aug 2021 & 19,800 & 10.0 & 16,938 & 85.5 \\
        \hline
        \textbf{Total} & & & 33,183 & -- & 28,342 & -- \\
        \hline
    \end{tabular}
\end{table}

The data used in this analysis covers a total of 34 surveys conducted on the panel between November 2019 and April 2023. The response rates for these surveys, conditional on panel membership, ranged between 32\% and 68\%, with a mean of 50\% . Approximately 90\% of the surveys were completed online, with the remainder completed by phone. Phone interviews were used for panel members who do not have access to the internet and for initial online nonrespondents. In early surveys all nonrespondents to the online survey, for whom a telephone number was available, were approached to be interviewed by phone. In later waves, telephone invitations were targeted at groups least likely to respond. Across surveys, approximately (10\%) of interviews were conducted on the phone and around (85\%) of these were online nonresponders, the remaining (15\%) being offliners. The panel is periodically purged of panelists who fail data quality checks and are persistent nonresponders (defined as six consecutive failures to respond). 

The initial dataset comprises the 27,127 panel members who were invited to participate in at least one panel survey. Of them, those who received only one invitation were used for just one part of the results, the estimated baseline hazard of first survey nonresponse. Our main analysis, where we  examine the influence of previous survey experience and personality on subsequent survey response, was then restricted to the subset of 11,976 panel members who completed at least one survey and were invited to complete a second. For them, we used data on all the survey invitations they received up to their last observed invitation or the first one that they failed to respond to, whichever was earlier.  

The number of survey invitations for this main analysis sample varies between 2 and 23, with the largest proportion of panel members (36\%) with two only. The mean number of survey invitations was 4.5, and the median was 3. The distribution of responses at each survey invitation used in the analysis is shown in Table A1 in the Appendix. The number of days between two successive invitations varied between 4 and 846 days, with a mean of 90 and a median of 68. The interval between recruitment and the first survey invitation varied from 8 to 564 days (mean = 126, median = 84). 

\section{Methods: Discrete-time survival analysis}
We use survival analysis to investigate how features of earlier survey experiences are related to time to first nonresponse. Here “time” is the sequential number of survey invitations a panel member received after completing the initial recruitment survey, coded as 1, 2, 3,…. Time 1 corresponds to the first time they were invited to take part in a survey, time 2 is the second time they were invited, and so on. In the discussion of the analysis, times will be referred to interchangeably as survey invitations. Actual calendar time between invitations, in days, is also included as an explanatory variable in the analysis.

The event of interest is the first time a panelist did not complete a survey they were invited to take. Because the invitation time is a discrete variable, methods of discrete-time survival analysis were used to model it. We use logistic models of the general form 

\begin{equation}
\text{logit}[P(Y_{it} =1 | \mathbf{X}_{it}; Y_{i,t'<t} =0)] = \alpha_t + \boldsymbol{\beta} \mathbf{X}_{it}
\label{logit}
\end{equation}
for survey invitation \( t \) for panel member \( i \). The response variable is the binary nonresponse indicator coded as \( Y_{it}=0 \) if the panelist responded to this invitation, and \( Y_{it}=1 \) if they did not. The conditioning on \( Y_{i,t'<t} =0 \) in (\ref{logit}) denotes that the probability is for nonresponse at time \( t \), given the panelist had responded to all the previous invitations up to time \( t-1 \). This probability is the \emph{hazard} of (first) nonresponse at time \( t \).          

On the right-hand side of (\ref{logit}), \( \mathbf{X}_{it} \) are predictors (explanatory variables), including the measures of survey experience up to time \( t \).  The \( \alpha_t \) are fixed effects for each time \( t \). The probabilities \( P(Y_{it} =1 | \mathbf{X}_{it} =0; Y_{i,t'<t} =0) = [1+\exp(-\alpha_t)]^{-1} \) give the hazard as a function of time when all the \( \mathbf{X}_{it} \) are zero, or the \emph{baseline hazard}. Since each time has its own fixed effect, the shape of the baseline hazard is unconstrained. The hazard is then moved up or down by different values of the explanatory variables. The \( \boldsymbol{\beta} \) are their coefficients, which  describe associations (as log odds ratios) between the explanatory variables and the hazard of first nonresponse. The variables in $\mathbf{X}_{it}$ can be time-varying (such as the survey experience measures) or constant over time (such as the personality scores of the respondent measured in the recruitment survey). They can also include interactions between explanatory variables and time $t$, to allow for this association to change as the number of invitations increases. 

The data are in person-time format, where each combination of \( i \) (panel member) and \( t \) (time) is a separate observation. Each person contributes observations up to time \( T_i \) which is the time of their first nonresponse or the last invitation observed for them, whichever is earlier. If the person replied even at time \( T_i \), their time to first nonresponse is ‘censored’ in the terminology of survival analysis. All the person-time observations can be treated as independent for purposes of maximum likelihood estimation, so this kind of discrete-time survival analysis can be done by fitting model (\ref{logit}) to the person-time data using standard procedures for binary logistic modelling. The analysis here was done using R, version 4.2.2 (R Core Team, 2022).

\section{Variables}
We use paradata to derive measures of \emph{objective respondent burden} and other aspects of previous survey experience to be used as the predictors in the models. There are two different measures of survey duration. The first is the individual-level response time (in minutes) for the most recent previous survey that the respondent completed. However, as a measure of burden, this is confounded with individual differences that influence response time such as age, education, topic interest, and so on \citep{sturgisInterviewerContributionVariability2021}. Therefore, we also include, as another measure of survey length, a measure of the average response time for all respondents who took that same survey. It is defined as the median time (in minutes) that these respondents took to complete the survey. Due to the sub-sampling approach implemented for OPP surveys, where a minority of panelists is invited to complete each survey, there is considerable variability in the duration of the last survey completed. This is an advantage of the OPP design over conventional panel surveys, were every panel member is invited to complete every wave, making the length of the previous survey the same for all respondents. There were some respondents with very long individual response times, defined beyond 1.5 times the IQR. These long response times are likely to result from exogenous factors, such as breaks and interruptions, rather than actual slow completions \citep{turnerCanResponseLatencies2015c}. We tried three ways of treating these cases, case-wise deletion, truncating them to 1.5 times the IQR and leaving them at their measured values. The results are substantively unchanged for each specification. Here we present the results under case-wise deletion.

We also include interactions with time (survey invitation) for both measures of survey duration to test whether longer surveys or longer response times have a different effect on subsequent response propensity as a respondent completes more surveys as a panel member. In addition to these survey duration variables, we include measures of the time (in days) between the recruitment survey and the first panel survey invitation and the time between the current invitation and the previous survey completion. It seems plausible that longer durations for both variables may be associated with a higher hazard of nonresponse. The first recruitment survey used mixed modes, with interviews conducted both face-to-face and online (push-to-web), so we also include a dummy indicator for the mode of the recruitment survey to assess whether this affects subsequent response propensity. 

Our indicator of \emph{subjective respondent burden} is a self-reported measure of survey enjoyment which was administered at the end of each survey. Respondents were asked, ‘We hope you have enjoyed doing our survey but let us know what you thought about it’ with response options 'I enjoyed all of it', 'I enjoyed some of it but not all of it', and 'I did not enjoy any of it'. Most respondents reported enjoying all of the survey (68\% on average over all surveys) with 30\% reporting they enjoyed some but not all and just 2\% saying they didn't enjoy any of it. We recoded this item to a binary (non-enjoyment) indicator which was 0 if the respondent reported that they enjoyed all of the most recent previous survey that they completed, and 1 otherwise (i.e.\ if they failed to enjoy some or all of it). We included the interaction of this indicator and time, to test whether not enjoying a survey is more important earlier or later in the panel experience. We also included a summary measure of enjoyment of the surveys that the respondent had taken \emph{before} the most recent one. It is possible that survey enjoyment has a cumulative effect where their whole previous experience with the panel may affect subsequent response propensity \citep{kaplanPrefaceOverviewSpecial2022}. To test this possibility, we used the proportion of all surveys taken before the most recent one that were (not) enjoyed. 

As predictor variables on \emph{personality}, we  include short-form measures of the Big Five personality traits that were included in the recruitment survey. We also include their interaction with time. 

As \emph{control variables} we included other potential predictors of nonresponse, namely the respondents' age, sex, education, marital status, housing tenture, ethnic group, disability status, employment status, political engagement, region, UK born, and deprivation status of their local area. The Verian Public Voice OPP samples panelists stratified by estimated response propensity \citep{williamsHowWeBuilt2022}, with a higher selection probability for low response propensity groups. Our control variables are also correlated with the inclusion probabilities, so including them mitigates the possibility that the model estimates will be biased by over-sampling of low response propensity panelists. To check whether they sufficiently account for this oversampling, we also included the recruitment survey weight in the model \citep{kimWeightingSurveyAnalysis2013}. It was found to be non-significant. 


There was item missing data for the analysis sample of 11,976 panel members. A small amount of this was in the demographic variables included in the model. As this was only a small number of cases, we deleted these cases listwise, leaving a sample of 11,169 respondents and 42,036 person*time observations. The individual survey length variable was missing for all respondents to survey 19 due to a data error, amounting to 3421 person*time observations. As the cause of this missingness is effectively random, we also deleted these observations listwise.

Missing data in the survey enjoyment variable was treated differently. Across the 34 distinct surveys included in the analysis, this variable was missing for some or all of the respondents in 15 of the surveys, amounting to 10,295 person*time observations. To avoid the substantial loss of information and to mitigate potential bias, missing values for the survey enjoyment variable were imputed using methods of multiple imputation \citep{rubinMultipleImputationNonresponse1987}. 

For the imputation model, binary logistic random effects models were fitted with the binary non-enjoyment variable as the outcome, including random intercepts for the survey respondents and using the data for each respondent up to and including the survey invitation before their first nonresponse. Explanatory variables in this imputation model included the other explanatory variables in our analysis model and a dummy variable for whether or not the respondent responded to their next survey invitation. Also included were dummy variables (fixed effects) for each survey where there was partial nonresponse in the enjoyment variable, to allow for different estimated average levels of enjoyment in these specific surveys. This imputation model was fitted in Stata using the \texttt{melogit} function \citep{statacorpStataStatisticalSoftware2023} and estimated probabilities of survey enjoyment were calculated from it for each respondent-survey observation with a missing value. Fifty sets of imputed values for them were drawn as Bernoulli random variables with these probabilities, to create fifty completed datasets. Our main analysis models (i.e.\ the survival analysis models for first nonresponse) were then fitted to each of these datasets, and results for them were combined using standard multiple-imputation methods to obtain the final estimates of interest. This last step was performed using the \texttt{mice} package in R \citep{vanbuurenMiceMultivariateImputation2011}.

\section{Results}
\begin{figure}[t]
    \centering
    \includegraphics[scale=0.9]{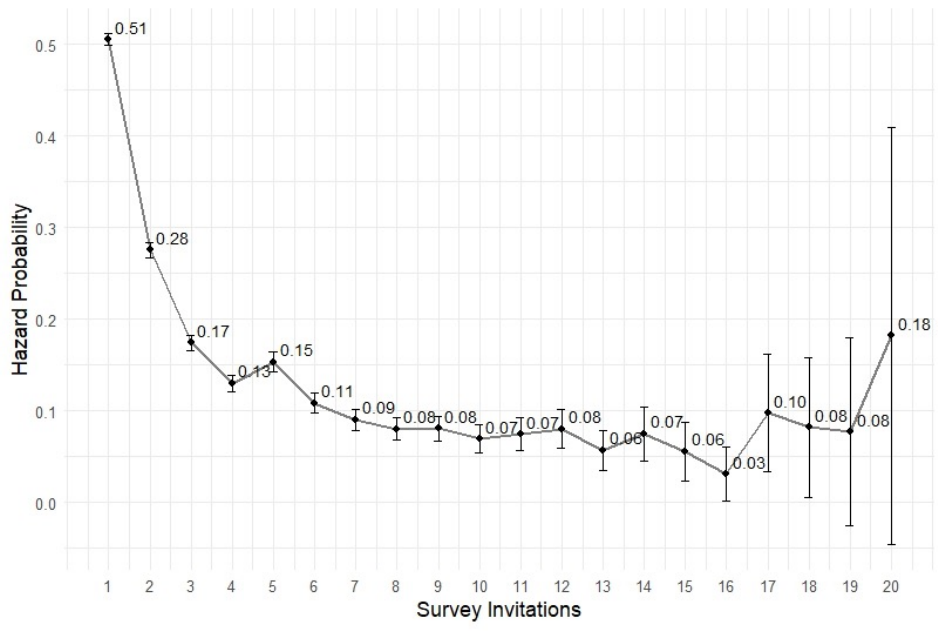}
    \caption{Overall hazard of first nonresponse (with 95\% confidence intervals), estimated from a model without explanatory variables.}
    \label{fig:baseline_hazard}
\end{figure}

We present first results from the discrete-time survival model (\ref{logit}) without any explanatory variables $\mathbf{X}_{it}$, i.e.\ including only the invitation fixed effects $\alpha_t$. For this baseline model, we also include the data on the response to the first survey invitation (time $t=1$ in our notation), irrespective of whether a respondent ever received a second one. This shows the overall pattern of first nonresponse among all panel members who were ever sampled for a survey. Figure~\ref{fig:baseline_hazard} presents the estimated hazards from this model with their (95\%) confidence intervals. For invitation 1 this is simply the proportion (51\%) of panel members who did not answer the first survey they were invited to. After that the hazard is the probability of nonresponse at a given invitation, conditional on having responded to all previous invitations. For example, of those who did reply to their first invitation, an estimated 28\% do not reply to their second, and so on. It is clear that panel members are at substantially greater risk of nonresponse for the first two invitations. From the third invitation onward the hazard becomes smaller and eventually stabilizes at less than 10\% for each additional invitation (after around invitation 16 the remaining samples of panel members with that many invitations are small and the estimates are very imprecise). Clearly, then, the first invitations on an OPP present the greatest opportunity for implementing interventions aimed at preventing nonresponse and attrition from the panel.

Tables \ref{t_models_1} and \ref{t_models_2} present the results for the main models, incorporating the explanatory and control variables. We report estimates from a model containing main effects only and from a model which also includes interactions with time (number of invitations). The sample for these models is now restricted to panel members with at least one previous survey completion. For the sake of parsimony of display, these tables show estimates for the main explanatory variables only, omitting the control variables and time fixed effects (Table A3 in the Appendix shows the full sets of estimates for these models). Considering first the model with main effects only, the coefficients in Tables \ref{t_models_1} and \ref{t_models_2} are generally statistically significant and in the expected direction. Completing the previous survey online is strongly associated with a lower hazard of nonresponse, which makes sense when it is recalled that the phone mode is used primarily as a means of converting initial online nonrespondents to interviews. A phone interview therefore likely acts as an indicator of a latent disposition toward nonresponse. The mode of the recruitment survey, however, does not appear to influence panel response propensity, with panelists who were recruited via push-to-web having a similar hazard as those recruited face-to-face. 

As noted earlier, the frequency of invitations, and hence the time between them, varies substantially between panel members. The estimates show that the size of this gap is consequential for nonresponse; the longer the gap since the previous invitation, the higher the risk of nonresponse. Any invitation gap greater than 30 days increases the hazard of nonresponse but the risk is greatest when the previous survey was completed over 112 days ago. This is likely because less frequent contact with panelists leads to lower engagement with the OPP with engagement itself predictive of nonresponse. 

A similar effect is observed for the number of days between the recruitment survey and first survey invitation, although this is less incremental, with the hazard of nonresponse only increasing at over 100 days and no further increase evident beyond that. The estimates for both these duration variables suggest that survey designers can mitigate nonresponse to OPP surveys by not delaying the first invitation too long after the recruitment survey and by issuing sufficiently frequent invitations thereafter, thereby reducing the gap between invitations (although clearly this must be balanced against the number of surveys that need to be administered commercially). 

There are two variables that index the duration of the previous interview, the individual interview length and the median time taken to complete the previous survey across all respondents who took it. Interestingly, both are highly significant but in opposite directions, controlling for the other; the longer someone takes to complete an interview, the less likely they are to be a norespondent at the next invitation --- but the longer the average duration of their previous interview, the more likely the panelist will be to fail to provide an interview at the next invitation. This difference reflects the fact that the individual interview duration is affected by a host of individual characteristics in addition to the number and types of questions that need to be answered. For the average duration measure, on the other hand, most of these individual differences cancel out, making it a more appropriate measure of burden arising from the length of the questionnaire. The individual response time, by comparison, likely represents the level of engagement the respondent has in the survey, with more engaged respondents taking longer to complete, \textit{ceteris paribus}. 

In terms of subjective respondent burden, a panelist who reported that they did not enjoy all of the previous survey has a significantly higher hazard of nonresponse at the next invitation. The measure of the proportion of earlier surveys the respondent did not enjoy was non-significant, indicating that there is no 'memory' from previous assessments of survey enjoyment, once the most recent experience is accounted for. 

The final substantive variables are the Big Five personality dimensions. Consistent with the existing literature, we find large and significant effects for some but not all of these factors. People who score high on Agreeableness and Extraversion are more likely to be nonrespondents across all survey invitations, while more Conscientious panelists are more likely to respond. This is consistent with a mechanism whereby Conscientious individuals feel a stronger sense of personal obligation to complete surveys once they had previously agreed to do so, while Extraverts and people who are high on Agreeableness are less engaged with what can be a boring task. Openness and Negative emotionality are not significantly related to nonresponse in the main effects model. 

\begin{sidewaystable}
\centering
\renewcommand{\arraystretch}{1.3}
\begin{minipage}{0.9\textwidth} 
\caption{Discrete-time survival models where outcome is first nonresponse}
\label{t_models_1}

\vspace{0.5em} 
\begin{tabular}{p{9cm} p{2cm} p{2cm} p{2cm} p{2cm}}
  \toprule
  & \multicolumn{2}{c}{\textbf{Main Effects}} & \multicolumn{2}{c}{\textbf{Interactions}} \\
  \textbf{Variable} & \textbf{B} & \textbf{SE} & \textbf{B} & \textbf{SE} \\
  \midrule
  Previous mode (ref: Online) & 1.264*** & 0.066 & 1.056*** & 0.121 \\
  Previous individual survey length (min) & -0.013*** & 0.002 & -0.014*** & 0.002 \\
  Previous average survey length (min) & 0.019*** & 0.003 & 0.006 & 0.005 \\
  \multicolumn{5}{l}{\emph{Days since last invitation (ref: $<$30 days)}} \\
  30--68 days & 0.104* & 0.044 & 0.090* & 0.044 \\
  69--112 days & 0.157*** & 0.045 & 0.151*** & 0.045 \\
  Over 112 days & 0.320*** & 0.046 & 0.309*** & 0.046 \\
  Did not enjoy previous survey & 0.315*** & 0.038 & 0.333*** & 0.066 \\
  Proportion earlier surveys not enjoyed & 0.050 & 0.061 & 0.054 & 0.062 \\
  \multicolumn{5}{l}{\emph{Recruitment to 1st invitation (ref: $<$66 days)}} \\
  66--99 days & -0.028 & 0.040 & -0.035 & 0.040 \\
  100--241 days & 0.221*** & 0.043 & 0.213*** & 0.044 \\
  Over 241 days & 0.179*** & 0.053 & 0.167** & 0.053 \\
  Push-to-web recruitment (ref: Face-to-face) & -0.107. & 0.059 & -0.113. & 0.059 \\
  \bottomrule
\end{tabular}

\vspace{0.8em}
\footnotesize
Note: Survey invitation coefficients, intercept, and demographics are shown in Table A4 (Appendix). \\* $p<0.05$, ** $p<0.01$, *** $p<0.001$
\end{minipage}

\end{sidewaystable}

\begin{sidewaystable}
\centering
\renewcommand{\arraystretch}{1.3}
\begin{minipage}{0.9\textwidth} 
\caption{Discrete-time survival models where outcome is first nonresponse – Continued.}
\label{t_models_2}

\vspace{0.5em} 
\begin{tabular}{p{9cm} p{2cm} p{2cm} p{2cm} p{2cm}}
  \toprule
  & \multicolumn{2}{c}{\textbf{Main Effects}} & \multicolumn{2}{c}{\textbf{Interactions}} \\
  \textbf{Variable} & \textbf{B} & \textbf{SE} & \textbf{B} & \textbf{SE} \\
  \midrule
  \multicolumn{5}{l}{\emph{Personality traits}} \\
  Agreeableness & 0.083*** & 0.020 & 0.156*** & 0.036 \\
  Conscientiousness & -0.095*** & 0.020 & -0.109** & 0.036 \\
  Extraversion & 0.062*** & 0.018 & 0.094** & 0.033 \\
  Negative emotionality & -0.030. & 0.016 & -0.019 & 0.027 \\
  Open-mindedness & -0.005 & 0.020 & -0.068* & 0.035 \\
  \multicolumn{5}{l}{\emph{Interactions with time}} \\
  Did not enjoy previous survey $\times$ No. of invitations & -- & -- & -0.005 & 0.014 \\
  Previous phone survey $\times$ time & -- & -- & 0.070* & 0.034 \\
  Previous avg. survey length $\times$ time & -- & -- & 0.003*** & 0.001 \\
  Agreeableness $\times$ time & -- & -- & -0.019* & 0.008 \\
  Conscientiousness $\times$ time & -- & -- & 0.004 & 0.007 \\
  Extraversion $\times$ time & -- & -- & -0.008 & 0.007 \\
  Negative emotionality $\times$ time & -- & -- & -0.003 & 0.006 \\
  Open-mindedness $\times$ time & -- & -- & 0.016* & 0.007 \\
  \midrule
  N & \multicolumn{4}{c}{27,127} \\
  \bottomrule
\end{tabular}

\vspace{0.8em}
\footnotesize
Note: Survey invitation coefficients, intercept, and demographics are shown in Table A4 (Appendix). \\* $p<0.05$, ** $p<0.01$, *** $p<0.001$
\end{minipage}

\end{sidewaystable}

\clearpage

\begin{figure}[t!] 
    \centering
    \includegraphics[scale=0.9]{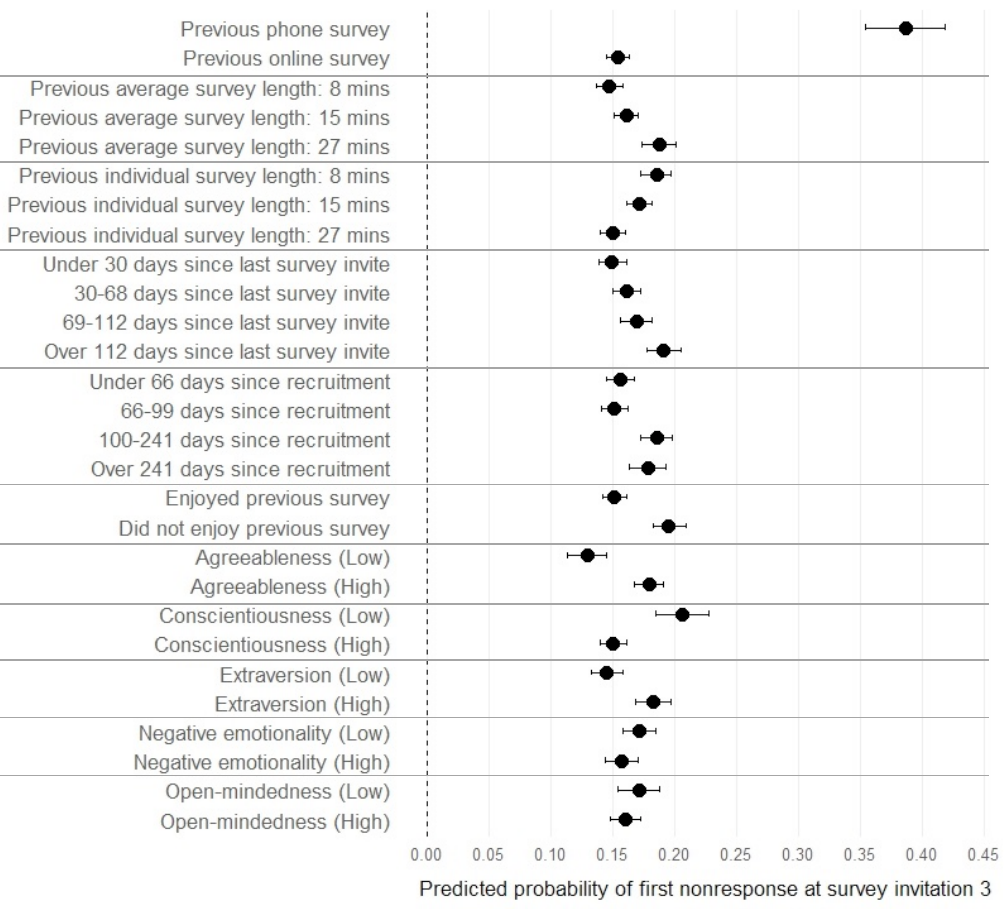}
    \caption{Predicted probability (with 95\% confidence interval) of first nonresponse at invitation 3, given values of individual predictor variables and averaged over the sample distribution of the other predictors.}
    \label{f_probs1}
\end{figure}

A better sense of the substantive magnitude of these estimates can be obtained from the marginal fitted probabilities presented in Figure \ref{f_probs1}, and comparisons between them. For each predictor in the interactions model, fitted hazards (probabilities of first nonresponse) were calculated with all other variables set at their observed values and then averaged across all observations to obtain the marginal probabilities. We used for this the sample of data on observations at invitation 3, i.e.\ those who had received and responded to their first two invitations.  

The strongest predictor of nonresponse is, for the reasons discussed above, the survey mode. A panelist interviewed for the previous survey by phone has a nonresponse probability of 38\% for the third invitation compared to just 16\% for those interviewed online. 

More interestingly, enjoyment of the survey is also very consequential, with the probability of nonresponse increasing by 33\% (from 15\% to 20\%) for a panelist who reports that they did not enjoy some of the previous survey. The probability of nonresponse also increases substantially when the median length of the previous survey across all who took it is long (27 minutes) compared to short (8 minutes). In contrast, a short individual response time for the previous survey (8 minutes) has a predicted probability of 19\% compared to 15\% when the individual response time was long (27 minutes). For the time between invites, the predicted probability increases from 15\% when the last survey was under 30 days previously to 19\% when it was more than 112 days ago. Inviting a new panelist to complete a survey soon after recruitment is also important, with the probability increasing from around 16\% when the first invite is within a hundred days to around 18\% when it is longer than that. 

Respondent personality also has large associations with nonresponse. The comparisons in Figure \ref{f_probs1} are between those scoring highest (5) against those scoring lowest (1) on the scale of each Big Five dimension that was used in the recruitment survey. While Negative emotionality and Open-mindedness are only weakly related to failure to respond at the next invitation, the remaining dimensions of the Big Five exhibit substantial effects. The least Conscientious panel members have a nonresponse probability of 21\% compared to 15\% for the most. A similar magnitude, though in the opposite direction, is observed for Agreeableness (13\%) compared to 17\%, while those scoring lowest on Extraversion are around 2 percentage points less likely to be nonrespondents (13\%) than those scoring highest (15\%). It is notable that these effects for Conscientiousness and Agreeableness are of equivalent magnitude to the survey enjoyment variable.

\clearpage 

\begin{figure}[htbp]
    \centering
    \includegraphics[scale=0.6]{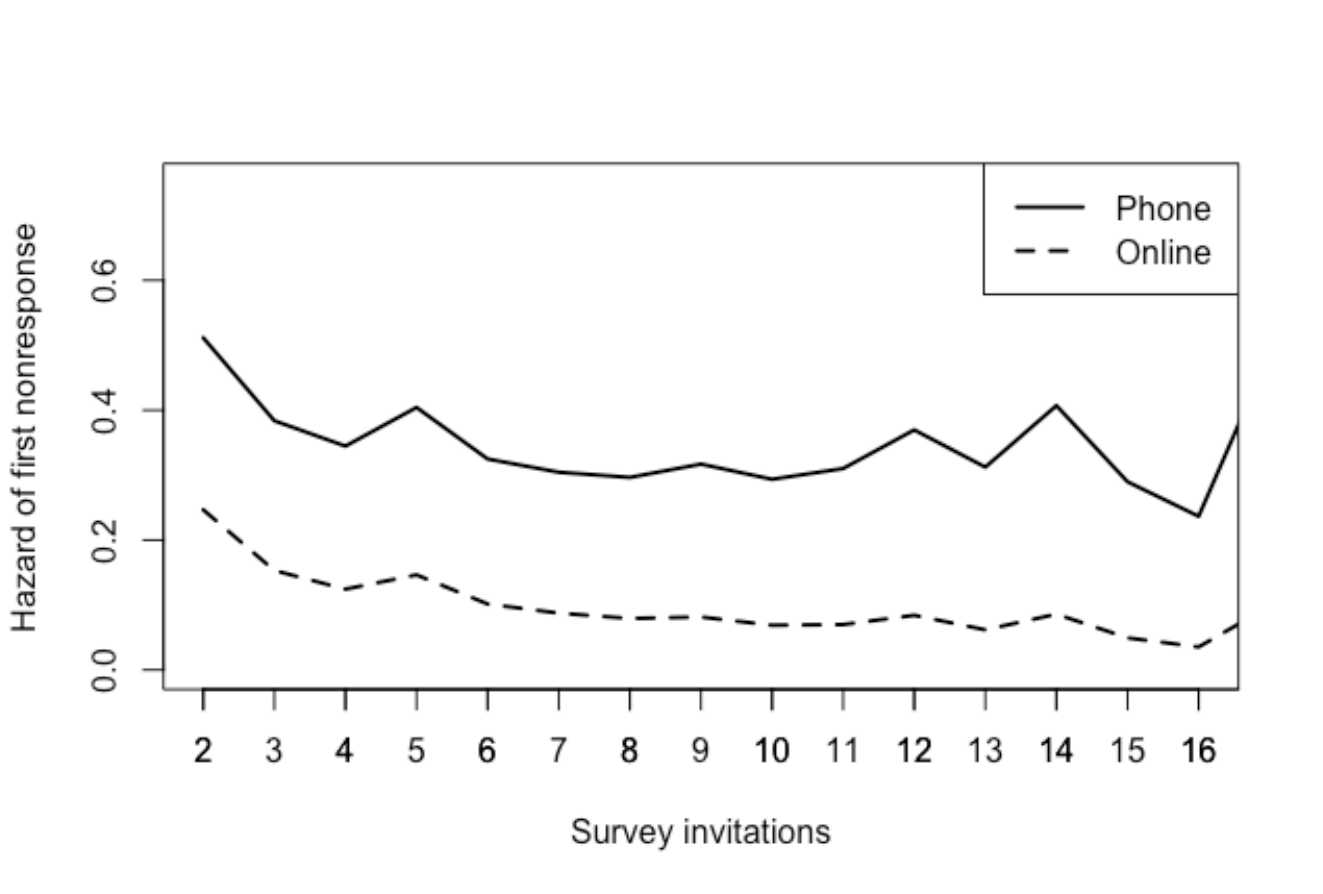}
    \caption{Predicted probability of first nonresponse at different survey invitations, given the mode of most recent previous survey invitation. These probabilities were calculated based on a model which included an interaction between time (survey invitation) and mode.}
    \label{f_mode_by_time}
\end{figure}

An advantage of the survival model is that we can assess whether the effect of survey experience and burden variables change as panelists take more surveys over time. This is done by including in the model an interaction between time and these variables. There are good reasons to expect that such interactions might be evident, as the nature of these experiences may be partly cumulative. For example, a long or unenjoyable survey may be more likely to result in nonresponse when a panelist is new compared to when they have completed a large number of surveys. Additionally, of course, the composition of panelists at first invitation will be quite different from those who remain after ten surveys, with less engaged and motivated panelists dropping out earlier. Tables \ref{t_models_1} and \ref{t_models_2} provide some support for such expectations, with the difference in response propensity between phone and online modes and the average length of the previous survey increasing over survey invitations. No such effect is observed for survey enjoyment, however, the effect of which is constant over survey invitations. Two of the Big Five personality dimensions have significant interactions with time; more Agreeable people are more likely to be nonrespondents but this effect is stronger earlier in the panel, while the opposite is true of Open-mindedness. These time effects are not large, however. By way of illustration, the marginal probabilities of first nonresponse over time given for phone and online self-completion of the previous survey are shown in Figure \ref{f_mode_by_time}. Despite the presence of the interaction, any change over time in the difference between the two modes is minimal. 

We can also use the model estimates to compare hazards for different 
profiles of panelists, as shown in Table \ref{t_scenarios}. For example, if the previous survey was completed online, those who enjoyed completing a comparatively long survey less than 30 days ago and took longer than average to complete it, have a marginal predicted probability of first nonresponse at the third invitation of just 12\%. By way of contrast, this probability is more than twice as high (27\%) for those who did not enjoy completing a long survey comparatively quickly over 112 days ago. Probabilities of these kinds could be of used in targeting interventions for the purpose of minimising nonresponse to a forthcoming survey. 

\begin{sidewaystable}
\centering
\renewcommand{\arraystretch}{1.3}
\begin{minipage}{0.95\textwidth}
\caption{Predicted probability of first nonresponse at invitation 3, by different respondents characteristics}
\label{t_scenarios}

\vspace{0.5em}
\begin{tabular}{>{\raggedright\arraybackslash}p{4.5cm} *{8}{>{\centering\arraybackslash}p{1.5cm}}}
\toprule
\multirow{3}{4.5cm}{\textbf{Previous survey length}} 
& \multicolumn{8}{c}{\textbf{Previous survey mode}} \\
\cmidrule(lr){2-9}
& \multicolumn{4}{c}{Online mode} & \multicolumn{4}{c}{Phone mode} \\
\cmidrule(lr){2-5} \cmidrule(lr){6-9}
& \multicolumn{2}{c}{Enjoyed} & \multicolumn{2}{c}{Did not enjoy} 
& \multicolumn{2}{c}{Enjoyed} & \multicolumn{2}{c}{Did not enjoy} \\
\cmidrule(lr){2-3} \cmidrule(lr){4-5} \cmidrule(lr){6-7} \cmidrule(lr){8-9}
& $<$30 days & $>$120 days & $<$30 days & $>$120 days & $<$30 days & $>$120 days & $<$30 days & $>$120 days \\
\midrule
Average length = 13-16 mins, Individual length = 26+ mins & 0.12 & 0.15 & 0.13 & 0.18 & 0.31 & 0.38 & 0.34 & 0.43 \\
Average length = 22+ mins, Individual length = 26 + mins & 0.12 & 0.15 & 0.15 & 0.20 & 0.32 & 0.38 & 0.39 & 0.46 \\
Average length = 22+ mins, Individual length = 13-17 mins & 0.15 & 0.20 & 0.21 & 0.27 & 0.39 & 0.47 & 0.49 & 0.56 \\
\bottomrule
\end{tabular}

\vspace{0.8em}
\footnotesize
Note: Table shows predicted marginal probabilities of first nonresponse at invitation 3, derived from the interaction model in Table 1. Effects vary by mode of previous survey, enjoyment of previous survey, time since last invitation, and survey length.
\end{minipage}
\end{sidewaystable}

\clearpage
\section{Discussion}
Online probability panels are taking on an increasingly important role in modern public opinion and survey research, offering a cost-effective, high-quality data collection infrastructure using probability sampling methods and limiting survey access to only those who are sampled. However, low response rates at recruitment combined with cross-sectional nonresponse and attrition mean these data collection vehicles are prone to potentially large nonresponse biases \citep{mcpheeDataQualityMetrics2022}. Additionally, a feature of OPPs is that panel members are sampled repeatedly to take part in cross-sectional surveys, meaning that aspects of the survey experience, over which researchers have some control, are likely to influence subsequent response decisions. Our aim in this study has, therefore, been to better understand the characteristics of survey experience that drive nonresponse in OPPs, so that targeted measures and interventions can be crafted to mitigate the impact of nonresponse on sample representativeness. 

In terms of \textit{when} nonresponse occurs, our survival analysis found that the first three survey invitations are when the probability of first nonresponse is substantially at its highest. This is perhaps unsurprising as these first invitations are sent to a larger and more diverse pool of panel members, including those who are weakly engaged with the panel. Nonetheless, this finding suggests that interventions targeted at the first few invitations are likely to offer the greatest potential for reducing OPP nonresponse. After the fifth survey completion, the subsequent hazard of nonresponse is essentially stable at around 8\% per survey invite. 

Regarding features of the survey experience, we found that survey mode is, by some margin, the strongest predictor of nonresponse, with those having been interviewed by phone at the previous survey having a nonresponse hazard of 38\% compared to just 16\% for those interviewed online. This striking difference is, however, essentially an artifact of the setting we considered. There are two factors that likely contribute to it. First, existing research has found that offliners are more prone to nonresponse than the online population \citep{blomDoesRecruitmentOffline2017, cornesseLongTermImpactDifferent2021,revillaWhatGainProbabilityBased2016} and phone is the mode offered to offliners in the Verian Public Voice OPP that we examined. Second, and more importantly, this OPP issues invitations to complete a phone interview to initially nonresponding online panel members and this group constitutes the great majority of all phone respondents. As initial nonrespondents, it is likely that this group has a higher propensity to nonresponse at this invitation compared to the average panel member. This implies that this mode effect is unlikely to be causal in nature but rather reflects the characteristics of the group that completes their interviews this way. That being so, there would be little benefit from implementing strategies that seek to shift panel members from phone to online self-completion. 

Our results show that objective and subjective measures of respondent burden have large and independent effects on subsequent response propensity. We evaluated the effect of a general summary measure of subjective burden, in which respondents to each survey are asked to report whether they enjoyed all, some, or none of the survey. Consistent with existing research on subjective indicators \citep{gummerNoteHowPrior2020,lugtigUsingParadataExplain2018,struminskayaDataQualityProbabilitybased2014} we found that not enjoying the survey increases the hazard of nonresponse at the next invitation by approximately five percentage points. It is possible that items tapping assessments of different dimensions of survey burden would be more effective in predicting response propensity and this would, we believe, be a useful direction for future research. Nonetheless, our research shows that this simple and intuitive question can serve as an effective way of tapping subjective survey burden that is strongly predictive of subsequent nonresponse. 

Regarding objective measures of burden, we considered two different measures of survey duration, a variable which has long been considered to drive down response \citep{kaplanPrefaceOverviewSpecial2022}. These were the individual respondent's response time to their most recent previous survey and the median response time to the previous survey across all the respondents who completed it. Interestingly, these measures both had significant associations with nonresponse but in opposite directions. Controlling for each other, longer average duration of the previous survey was associated with a higher hazard, but  longer individual interview length with a lower hazard, of subsequent nonresponse. We believe this goes some way to explaining the mixed evidence on survey length in the existing nonresponse literature. While the average duration of the previous survey is unambiguously a measure of the time-cost to respondents of completing the survey, the individual survey length is affected by a range of observed and unobserved respondent characteristics, in addition to the length of the questionnaire. We therefore argue that the individual response time effect we have observed reflects respondent engagement with the questionnaire, all else equal. Accordingly, using an unadjusted individual response time as a measure of burden is not appropriate and should be avoided.   

In addition to these survey duration variables, we also considered the effect of the lag between the recruitment survey and the first panel survey invitation, and the gap between the previous survey completed and the next invite. Longer durations for both variables would be expected to lead to higher nonresponse due to the loss of engagement with and commitment to the panel through inactivity. Our results support these expectations, with the hazard of first nonresponse increasing the longer the duration of both gaps. These findings suggest that sampling strategies should take these durations into account when specifying the design of any single OPP survey in order to minimise the number of panelists with long durations between times when they are invited to engage with the panel. 

The fact that the objective and subjective measures of burden had independent effects on subsequent response propensity demonstrates that the kind of summary subjective assessment of burden we have used here does not act as a complete mediator of the aspects of the survey experience that are thought to contribute to burden. Another way of saying this is that when respondents assess the extent to which they enjoyed the survey, they are not factoring in all aspects of the survey experience that are relevant to their response decision at the next invitation. Of course, this is just a single indicator of subjective burden, and it is possible that a more comprehensive battery of items would work better in this regard, though including more questions of this kind would itself increase the burden experienced by respondents, making the measurements partly endogenous. 

In terms of personality, we find substantively large effects in the range of 3-5 percentage points for those scoring at the top compared to the bottom on Agreeableness, Conscientiousness, and Extraversion. It is difficult to assess how well these findings align with the existing evidence on personality and nonresponse because, as discussed earlier, the existing evidence is itself quite mixed. We note that our result for Conscientiousness, where more Conscientious individuals are more likely to respond, is consistent with both theoretical expectations and existing findings. It is less clear, however, why more Agreeable and Extravert people should be more likely to be nonrespondents, albeit that both effects have been observed in existing studies. Resolving these kinds of inconsistencies cannot realistically be achieved through a single study and we expect that a fuller understanding of these personality effects will best be addressed through meta-analytic studies and systematic review. We hope that our findings will be useful in that regard. 

We find some evidence that how different aspects of the survey experience affect nonresponse is different at earlier than at later stages of an OPP, although these effects were generally rather small in magnitude. In additional analyses not reported here, we found no evidence that survey enjoyment and survey duration have different associations with nonresponse at different points in the panel life-cycle.  

Lastly, we note that our findings in this paper raise questions as well as answering others. First, our estimates have little to say about what aspects of survey experience determine survey enjoyment and other subjective measures of burden, we see this as an important question for researchers in this area. Second, our analysis focuses on only one response outcome - first nonresponse in the panel, and other outcomes may show different patterns over time and variable relationships with the survey experience and burden variables. For example, full attrition from the panel is something we would like to evaluate in developments of this research. We are also interested in the factors that lead a panel member to return to responding, after failing to respond to one or more invitations, as these will be useful in guiding panel retention strategies. Our findings here are not, then, the last word on this important area of nonresponse research but, rather, initial steps toward deepening what we know about the drivers of nonresponse in OPPs and helping to develop mitigation strategies, while also suggesting fruitful avenues for future research.

\section{Appendix}
\label{sec:appendix}

\begin{table}[H]
\centering
\renewcommand{\arraystretch}{1.3}

\vspace{1em}
\textbf{Table A1.} Survey response at each survey invitation
\vspace{1em}

\begin{tabular}{>{\raggedright\arraybackslash}p{4.5cm}ccc}
\toprule
\textbf{Survey invitation} & \textbf{Total cases} & \textbf{Responded} & \textbf{Did not respond (event)} \\
\midrule
1  & 27,127 & 13,419 & 13,708 \\
2  & 11,976 & 8,677  & 3,299  \\
3  & 7,649  & 6,317  & 1,332  \\
4  & 5,578  & 4,853  & 725    \\
5  & 4,248  & 3,597  & 651    \\
6  & 3,110  & 2,774  & 336    \\
7  & 2,373  & 2,159  & 214    \\
8  & 1,826  & 1,680  & 146    \\
9  & 1,432  & 1,317  & 115    \\
10 & 1,125  & 1,047  & 78     \\
11 & 847    & 784    & 63     \\
12 & 624    & 574    & 50     \\
13 & 440    & 415    & 25     \\
14 & 307    & 284    & 23     \\
15 & 197    & 186    & 11     \\
16 & 130    & 126    & 4      \\
17 & 82     & 74     & 8      \\
18 & 49     & 45     & 4      \\
19 & 26     & 24     & 2      \\
20 & 11     & 9      & 2      \\
21 & 4      & 4      & 0      \\
22 & 1      & 1      & 0      \\
23 & 1      & 1      & 0      \\
\bottomrule
\end{tabular}
\end{table}

\begin{table}[htbp]
\centering
\renewcommand{\arraystretch}{1.3}
\begin{minipage}{0.7\textwidth}
\textbf{Table A2: Time coefficients and intercept estimates from the baseline discrete-time survival model for time to first nonresponse (no covariates)}

\vspace{0.5em}
\begin{tabular}{p{7cm} p{2.5cm} p{2.5cm}}
\toprule
\textbf{Survey invitation} & \textbf{B} & \textbf{SE} \\
\midrule
\textbf{Survey invitations (ref: Survey invitation 1)} \\
Survey invitation 2  & -0.988*** & 0.024 \\
Survey invitation 3  & -1.578*** & 0.033 \\
Survey invitation 4  & -1.922*** & 0.042 \\
Survey invitation 5  & -1.731*** & 0.044 \\
Survey invitation 6  & -2.132*** & 0.059 \\
Survey invitation 7  & -2.333*** & 0.073 \\
Survey invitation 8  & -2.464*** & 0.087 \\
Survey invitation 9  & -2.459*** & 0.098 \\
Survey invitation 10 & -2.618*** & 0.118 \\
Survey invitation 11 & -2.543*** & 0.132 \\
Survey invitation 12 & -2.462*** & 0.148 \\
Survey invitation 13 & -2.831*** & 0.206 \\
Survey invitation 14 & -2.535*** & 0.217 \\
Survey invitation 15 & -2.849*** & 0.311 \\
Survey invitation 16 & -3.471*** & 0.508 \\
Survey invitation 17 & -2.246*** & 0.372 \\
Survey invitation 18 & -2.442*** & 0.522 \\
Survey invitation 19 & -2.506*** & 0.736 \\
Survey invitation 20 & -1.525.   & 0.782 \\
Survey invitation 21 & Undefined & Undefined \\
Survey invitation 22 & Undefined & Undefined \\
Survey invitation 23 & Undefined & Undefined \\
\midrule
Constant & 0.021. & 0.012 \\
N        & \multicolumn{2}{c}{27,127} \\
\bottomrule
\end{tabular}

\vspace{1em}
\footnotesize{\textit{Note}: * $p<0.05$, ** $p<0.01$, *** $p<0.001$}
\end{minipage}
\end{table}

\clearpage
\begin{sidewaystable}
\centering
\renewcommand{\arraystretch}{1.3}
\begin{minipage}{\linewidth}
\textbf{Table A3: Time coefficients, demographic estimates and intercept from the full discrete-time survival model for time to first nonresponse (with covariates)}

\vspace{0.5em}
\begin{tabular}{p{9cm} p{2cm} p{2cm} p{2cm} p{2cm}}
\toprule
& \multicolumn{2}{c}{\textbf{Main effects model}} & \multicolumn{2}{c}{\textbf{Interaction effects model}} \\
\textbf{Variable} & \textbf{B} & \textbf{SE} & \textbf{B} & \textbf{SE} \\
\midrule
(Intercept) & -0.977*** & 0.184 & -0.909*** & 0.212 \\
\textbf{Survey invitations (ref: Survey invitation 2)} \\
Survey invitation 3 & -0.544*** & 0.048 & -0.562*** & 0.072 \\
Survey invitation 4 & -0.780*** & 0.053 & -0.821*** & 0.120 \\
Survey invitation 5 & -0.580*** & 0.055 & -0.648*** & 0.172 \\
Survey invitation 6 & -1.002*** & 0.070 & -1.085*** & 0.228 \\
Survey invitation 7 & -1.169*** & 0.086 & -1.284*** & 0.286 \\
Survey invitation 8 & -1.227*** & 0.098 & -1.366*** & 0.343 \\
Survey invitation 9 & -1.234*** & 0.111 & -1.393*** & 0.401 \\
Survey invitation 10 & -1.418*** & 0.134 & -1.563*** & 0.457 \\
Survey invitation 11 & -1.397*** & 0.154 & -1.557** & 0.521 \\
Survey invitation 12 & -1.213*** & 0.170 & -1.388* & 0.578 \\
Survey invitation 13 & -1.451*** & 0.224 & -1.638* & 0.649 \\
Survey invitation 14 & -1.098*** & 0.223 & -1.278* & 0.699 \\
Survey invitation 15 & -1.745*** & 0.366 & -2.015* & 0.817 \\
Survey invitation 16 & -2.048*** & 0.512 & -2.350* & 0.936 \\
Survey invitation 17 & -0.867* & 0.377 & -1.219 & 0.912 \\
Survey invitation 18 & -1.209* & 0.525 & -1.567 & 1.034 \\
Survey invitation 19 & -1.132 & 0.742 & -1.435 & 1.228 \\
Survey invitation 20 & -0.152 & 0.798 & -0.440 & 1.305 \\
Survey invitation 21 & -10.149 & 98.158 & -10.533 & 95.232 \\
Survey invitation 22 & -10.427 & 196.968 & -10.438 & 196.971 \\
Survey invitation 23 & -10.816 & 196.968 & -11.836 & 196.972 \\
\bottomrule
\end{tabular}
\end{minipage}
\end{sidewaystable}

\begin{sidewaystable}
\centering
\renewcommand{\arraystretch}{1.3}
\begin{minipage}{\linewidth}
\textbf{Table A3 (Part 2): Survey experience, timing, and personality predictors from the full discrete-time survival model for time to first nonresponse (with covariates)}

\vspace{0.5em}
\begin{tabular}{p{9cm} p{2cm} p{2cm} p{2cm} p{2cm}}
\toprule
& \multicolumn{2}{c}{\textbf{Main effects model}} & \multicolumn{2}{c}{\textbf{Interaction effects model}} \\
\textbf{Variable} & \textbf{B} & \textbf{SE} & \textbf{B} & \textbf{SE} \\
\midrule
Previous phone survey (ref: Online) & 1.264*** & 0.066 & 1.056*** & 0.121 \\
Previous individual survey length (minutes) & -0.013*** & 0.002 & -0.014*** & 0.002 \\
Previous average survey length (minutes) & 0.019*** & 0.003 & 0.006 & 0.005 \\
\textbf{Days since last survey invitation (ref: Under 30 days)} \\
30–68 days & 0.104* & 0.044 & 0.090* & 0.044 \\
69–112 days & 0.157*** & 0.045 & 0.151*** & 0.045 \\
Over 112 days & 0.320*** & 0.046 & 0.309*** & 0.046 \\
Did not enjoy all of the previous survey & 0.315*** & 0.038 & 0.333*** & 0.066 \\
Proportion of earlier surveys did not enjoy & 0.050 & 0.061 & 0.054 & 0.062 \\
\textbf{Days between recruitment and first survey invitation (ref: Under 66 days)} \\
66–99 days & -0.028 & 0.040 & -0.035 & 0.040 \\
100–241 days & 0.221*** & 0.043 & 0.213*** & 0.044 \\
Over 241 days & 0.179*** & 0.053 & 0.167** & 0.053 \\
Push-to-web recruitment (ref: Face-to-face) & -0.107. & 0.059 & -0.113. & 0.059 \\
\textbf{Personality traits} \\
Agreeableness & 0.083*** & 0.020 & 0.156*** & 0.036 \\
Conscientiousness & -0.095*** & 0.020 & -0.109** & 0.036 \\
Extraversion & 0.062*** & 0.018 & 0.094** & 0.033 \\
Negative emotionality & -0.030. & 0.016 & -0.019 & 0.027 \\
Open-mindedness & -0.005 & 0.020 & -0.068* & 0.035 \\
\bottomrule
\end{tabular}
\end{minipage}
\end{sidewaystable}

\begin{sidewaystable}
\centering
\renewcommand{\arraystretch}{1.3}
\begin{minipage}{\linewidth}
\textbf{Table A3 (Part 3): Demographic and regional predictors from the full discrete-time survival model for time to first nonresponse (with covariates)}

\vspace{0.5em}
\begin{tabular}{p{9cm} p{2cm} p{2cm} p{2cm} p{2cm}}
\toprule
& \multicolumn{2}{c}{\textbf{Main effects model}} & \multicolumn{2}{c}{\textbf{Interaction effects model}} \\
\textbf{Variable} & \textbf{B} & \textbf{SE} & \textbf{B} & \textbf{SE} \\
\midrule
\textbf{Demographics} \\
Age 25–34 (ref: Under 25) & -0.246*** & 0.061 & -0.248*** & 0.061 \\
Age 35–44 & -0.291*** & 0.063 & -0.293*** & 0.063 \\
Age 45–54 & -0.277*** & 0.064 & -0.278*** & 0.064 \\
Age 55–64 & -0.247*** & 0.064 & -0.247*** & 0.064 \\
Age 65–74 & -0.238*** & 0.069 & -0.238*** & 0.069 \\
Age 75+ & 0.107 & 0.085 & 0.106 & 0.085 \\
Male & 0.001 & 0.033 & 0.001 & 0.033 \\
Disability & 0.030 & 0.035 & 0.025 & 0.035 \\
White & -0.076 & 0.053 & -0.078 & 0.053 \\
Born in the UK & -0.054 & 0.047 & -0.057 & 0.048 \\
Degree & -0.059* & 0.034 & -0.056* & 0.034 \\
Own home & -0.221*** & 0.036 & -0.224*** & 0.036 \\
In work & -0.002 & 0.039 & -0.001 & 0.039 \\
Living with someone as a couple & -0.012 & 0.034 & -0.012 & 0.034 \\
\textbf{Region (ref: London)} \\
North East & 0.029 & 0.087 & 0.033 & 0.087 \\
North West & -0.071 & 0.063 & -0.067 & 0.063 \\
Yorkshire & -0.057 & 0.069 & -0.053 & 0.069 \\
East Midlands & -0.130* & 0.070 & -0.130* & 0.070 \\
West Midlands & 0.001 & 0.068 & 0.001 & 0.068 \\
East of England & -0.099 & 0.065 & -0.100 & 0.065 \\

\bottomrule
\end{tabular}
\end{minipage}
\end{sidewaystable}

\begin{sidewaystable}
\centering
\renewcommand{\arraystretch}{1.3}
\begin{minipage}{\linewidth}
\textbf{Table A3 (Part 4): Deprivation, political interest, interaction terms, and sample size from the full discrete-time survival model for time to first nonresponse (with covariates)}

\vspace{0.5em}
\begin{tabular}{p{9cm} p{2cm} p{2cm} p{2cm} p{2cm}}
\toprule
& \multicolumn{2}{c}{\textbf{Main effects model}} & \multicolumn{2}{c}{\textbf{Interaction effects model}} \\
\textbf{Variable} & \textbf{B} & \textbf{SE} & \textbf{B} & \textbf{SE} \\
\midrule
South East & -0.074 & 0.059 & -0.074 & 0.059 \\
South West & -0.015 & 0.067 & -0.016 & 0.067 \\
Northern Ireland & 0.436*** & 0.114 & 0.437*** & 0.114 \\
Scotland & 0.167* & 0.069 & 0.174* & 0.069 \\
Wales & 0.220** & 0.083 & 0.224** & 0.083 \\
Index of multiple deprivation decile & -0.005 & 0.006 & -0.005 & 0.006 \\
Follows politics & 0.011 & 0.034 & 0.012 & 0.034 \\
\textbf{Interaction terms} \\
Did not enjoy previous survey * No. of invitations & — & — & -0.005 & 0.014 \\
Previous phone survey * No. of invitations & — & — & 0.070* & 0.034 \\
Previous avg. survey length * No. of invitations & — & — & 0.003*** & 0.001 \\
Agreeableness * No. of invitations & — & — & -0.019* & 0.008 \\
Conscientiousness * No. of invitations & — & — & 0.004 & 0.007 \\
Extraversion * No. of invitations & — & — & -0.008 & 0.007 \\
Negative emotionality * No. of invitations & — & — & -0.003 & 0.006 \\
Open-mindedness * No. of invitations & — & — & 0.016* & 0.007 \\
\midrule
N & \multicolumn{4}{c}{27,127} \\
\bottomrule
\end{tabular}

\vspace{1em}
\footnotesize{\textit{Note}: * p$<$0.05, ** p$<$0.01, *** p$<$0.001}
\end{minipage}
\end{sidewaystable}


\begin{thebibliography}{}

\bibitem[Blom et~al., 2016]{blomComparisonFourProbabilityBased2016}
Blom, A.~G., Bosnjak, M., Cornilleau, A., Cousteaux, A.-S., Das, M., Douhou, S., and Krieger, U. (2016).
\newblock A {{Comparison}} of {{Four Probability-Based Online}} and {{Mixed-Mode Panels}} in {{Europe}}.
\newblock {\em Social Science Computer Review}, 34(1):8--25.

\bibitem[Blom et~al., 2017]{blomDoesRecruitmentOffline2017}
Blom, A.~G., Herzing, J. M.~E., Cornesse, C., Sakshaug, J.~W., Krieger, U., and Bossert, D. (2017).
\newblock Does the {{Recruitment}} of {{Offline Households Increase}} the {{Sample Representativeness}} of {{Probability-Based Online Panels}}? {{Evidence From}} the {{German Internet Panel}}.
\newblock {\em Social Science Computer Review}, 35(4):498--520.

\bibitem[Bosch and Maslovskaya, 2023]{boschUtilityProbabilitybasedOnline2023}
Bosch, O.~J. and Maslovskaya, O. (2023).
\newblock The utility of probability-based online surveys: A literature review.

\bibitem[Bradburn, 1978]{bradburnRespondentBurden1978}
Bradburn, N. (1978).
\newblock Respondent burden.
\newblock {\em Proceedings of the Survey Research Methods Section of the American Statistical Association}, pages 35--40.

\bibitem[Callegaro et~al., 2014]{callegaroOnlinePanelResearch2014}
Callegaro, M., Baker, R., Bethlehem, J., G{\"o}ritz, A.~S., Krosnick, J.~A., and Lavrakas, P.~J. (2014).
\newblock Online panel research.
\newblock In {\em Online {{Panel Research}}}, chapter~1, pages 1--22. John Wiley \& Sons, Ltd.

\bibitem[Cheng et~al., 2020]{chengPersonalityPredictorUnit2020a}
Cheng, A., Zamarro, G., and Orriens, B. (2020).
\newblock Personality as a {{Predictor}} of {{Unit Nonresponse}} in an {{Internet Panel}}.
\newblock {\em Sociological Methods \& Research}, 49(3):672--698.

\bibitem[Cornesse and Schaurer, 2021]{cornesseLongTermImpactDifferent2021}
Cornesse, C. and Schaurer, I. (2021).
\newblock The {{Long-Term Impact}} of {{Different Offline Population Inclusion Strategies}} in {{Probability-Based Online Panels}}: {{Evidence From}} the {{German Internet Panel}} and the {{GESIS Panel}}.
\newblock {\em Social Science Computer Review}, 39(4):687--704.

\bibitem[Gummer and Daikeler, 2020]{gummerNoteHowPrior2020}
Gummer, T. and Daikeler, J. (2020).
\newblock A {{Note}} on {{How Prior Survey Experience With Self-Administered Panel Surveys Affects Attrition}} in {{Different Modes}}.
\newblock {\em Social Science Computer Review}, 38(4):490--498.

\bibitem[Hansson et~al., 2018]{hanssonCanPersonalityPredict2018a}
Hansson, I., Berg, A.~I., and Thorvaldsson, V. (2018).
\newblock Can personality predict longitudinal study attrition? {{Evidence}} from a population-based sample of older adults.
\newblock {\em Journal of Research in Personality}, 77:133--136.

\bibitem[Jessop, 2017]{jessopDevelopingNatCenPanel2017}
Jessop, C. (2017).
\newblock Developing the {{NatCen Panel}}, {{August}} 2015 -- {{July}} 2017.
\newblock Technical report, NatCen.

\bibitem[Jessop and Williams, 2021]{jessopComparingFacetofaceOnline2021}
Jessop, C. and Williams, J. (2021).
\newblock Comparing face-to-face and online recruitment approaches: Evidence from two probability-based panels in the {{UK}}.
\newblock In {\em Non-Response in Longitudinal Studies}.

\bibitem[Jin and Kapteyn, 2022]{jinRelationshipSurveyBurden2022}
Jin, H. and Kapteyn, A. (2022).
\newblock Relationship {{Between Past Survey Burden}} and {{Response Probability}} to a {{New Survey}} in a {{Probability-Based Online Panel}}.
\newblock {\em Journal of Official Statistics}, 38(4):1051--1067.

\bibitem[Kaczmirek et~al., 2019]{kaczmirekBuildingProbabilitybasedOnline2019}
Kaczmirek, L., Phillips, B., Pennay, {\relax DW}., Lavrakas, {\relax PJ}., and Neiger, D. (2019).
\newblock Building a probability-based online panel: {{Life}} in {{Australia}}.

\bibitem[{Kantar Public}, 2021]{kantarpublicCommunityLifeSurvey2021}
{Kantar Public} (2021).
\newblock Community {{Life Survey Technical Report}} 2020/21.
\newblock Technical report, {Department for Digital, Culture, Media and Sport}.

\bibitem[Kaplan et~al., 2022]{kaplanPrefaceOverviewSpecial2022}
Kaplan, R.~L., Holzberg, J., Eckman, S., and Giesen, D. (2022).
\newblock Preface {{Overview}} of the {{Special Issue}} on {{Respondent Burden}}.
\newblock {\em Journal of Official Statistics}, 38(4):929--938.

\bibitem[Kennedy et~al., 2020]{kennedyAssessingRisksOnline2020}
Kennedy, C., Hatley, N., Lau, A., Mercer, A., Keeter, S., Ferno, J., and {Asare-Marfo}, D. (2020).
\newblock Assessing the {{Risks}} to {{Online Polls From Bogus Respondents}}.
\newblock Technical report, Pew Research Centre.

\bibitem[Kim and Skinner, 2013]{kimWeightingSurveyAnalysis2013}
Kim, J.~K. and Skinner, C.~J. (2013).
\newblock Weighting in survey analysis under informative sampling.
\newblock {\em Biometrika}, 100(2):385--398.

\bibitem[Kocar and Biddle, 2023]{kocarPowerOnlinePanel2023}
Kocar, S. and Biddle, N. (2023).
\newblock The power of online panel paradata to predict unit nonresponse and voluntary attrition in a longitudinal design.
\newblock {\em Quality \& Quantity}, 57(2):1055--1078.

\bibitem[Lugtig and Blom, 2018]{lugtigUsingParadataExplain2018}
Lugtig, P. and Blom, A. (2018).
\newblock Using paradata to explain attrition in the {{German Internet Panel}}.
\newblock In {\em Methodology of {{Longitudinal Surveys Conference}}}. United Kingdom, Essex.

\bibitem[Lugtig et~al., 2014]{lugtigNonresponseAttritionProbabilitybased2014a}
Lugtig, P., Das, M., and Scherpenzeel, A. (2014).
\newblock Nonresponse and attrition in a probability-based online panel for the general population.
\newblock In {\em Online {{Panel Research}}}, chapter~6, pages 135--153. John Wiley \& Sons, Ltd.

\bibitem[Lynn, 2014]{lynnLongerInterviewsMay2014}
Lynn, P. (2014).
\newblock Longer {{Interviews May Not Affect Subsequent Survey Participation Propensity}}.
\newblock {\em Public Opinion Quarterly}, 78(2):500--509.

\bibitem[McCrae and Costa~Jr., 1999]{mccraeFiveFactorTheoryPersonality1999}
McCrae, R.~R. and Costa~Jr., P.~T. (1999).
\newblock A {{Five-Factor}} theory of personality.
\newblock In {\em Handbook of Personality: {{Theory}} and Research, 2nd Ed}, pages 139--153. Guilford Press, New York, NY, US.

\bibitem[McPhee et~al., 2022]{mcpheeDataQualityMetrics2022}
McPhee, C., Barlas, F., Brigham, N., Darling, J., Dutwin, D., Jackson, C., Jackson, M., Krizinger, A., Little, R., Lorenz, E., Marlar, J., Mercer, A., Scanlon, P., Weiss, S., and Wronski, L. (2022).
\newblock Data quality metrics for online samples: {{Considerations}} for study design and analysis.
\newblock Technical report, American Association for Public Opinion Research.

\bibitem[Mercer and Lau, 2023]{mercerComparingTwoTypes2023}
Mercer, A. and Lau, A. (2023).
\newblock Comparing {{Two Types}} of {{Online Survey Samples}}.
\newblock Technical report, Pew Research Centre.

\bibitem[Minderop, 2023]{minderopPredictingPanelAttrition2023}
Minderop, I. (2023).
\newblock Predicting panel attrition using multiple operationalisations of response time.
\newblock {\em Survey Methods: Insights from the Field}.

\bibitem[Revilla et~al., 2016]{revillaWhatGainProbabilityBased2016}
Revilla, M., Cornilleau, A., Cousteaux, A.-S., Legleye, S., and {de Pedraza}, P. (2016).
\newblock What {{Is}} the {{Gain}} in a {{Probability-Based Online Panel}} of {{Providing Internet Access}} to {{Sampling Units Who Previously Had No Access}}?
\newblock {\em Social Science Computer Review}, 34(4):479--496.

\bibitem[Richter et~al., 2014]{richterPersonalityHasMinor2014}
Richter, D., K{\"o}rtner, J.~L., and Sa{\ss}enroth, D. (2014).
\newblock Personality has minor effects on panel attrition.
\newblock {\em Journal of Research in Personality}, 53:31--35.

\bibitem[Rosche et~al., 2020]{roscheSurveyAttitudeIndicator2020}
Rosche, B., Hox, J., and {de Leeuw}, E. (2020).
\newblock Survey attitude as indicator for survey climate and as predictor of nonresponse and attrition in a probability-based online panel.

\bibitem[Ro{\ss}mann and Gummer, 2015]{rossmannUsingParadataPredict2015}
Ro{\ss}mann, J. and Gummer, T. (2015).
\newblock Using {{Paradata}} to {{Predict}} and {{Correct}} for {{Panel Attrition}}.
\newblock {\em Social Science Computer Review}, 34.

\bibitem[Rubin, 1987]{rubinMultipleImputationNonresponse1987}
Rubin, D.~B. (1987).
\newblock {\em Multiple {{Imputation}} for {{Nonresponse}} in {{Surveys}}}.
\newblock Wiley {{Series}} in {{Probability}} and {{Statistics}}. Wiley, 1 edition.

\bibitem[Salthouse, 2014]{salthouseSelectivityAttritionLongitudinal2014}
Salthouse, T.~A. (2014).
\newblock Selectivity of {{Attrition}} in {{Longitudinal Studies}} of {{Cognitive Functioning}}.
\newblock {\em The Journals of Gerontology Series B: Psychological Sciences and Social Sciences}, 69(4):567--574.

\bibitem[Sa{\ss}enroth, 2013]{sassenrothImpactPersonalityParticipation2013}
Sa{\ss}enroth, D. (2013).
\newblock {\em The {{Impact}} of {{Personality}} on {{Participation Decisions}} in {{Surveys}}: {{A Contribution}} to the {{Discussion}} on {{Unit Nonresponse}}}.
\newblock Springer Fachmedien Wiesbaden GmbH, Wiesbaden, 1st ed edition.

\bibitem[Satherley et~al., 2015]{satherleyDemographicPsychologicalPredictors2015a}
Satherley, N., Milojev, P., Greaves, L.~M., Huang, Y., Osborne, D., Bulbulia, J., and Sibley, C.~G. (2015).
\newblock Demographic and {{Psychological Predictors}} of {{Panel Attrition}}: {{Evidence}} from the {{New Zealand Attitudes}} and {{Values Study}}.
\newblock {\em PLOS ONE}, 10(3):e0121950.

\bibitem[Schwerdtfeger and Bach, 2024]{schwerdtfegerHowDoesBroadband2024}
Schwerdtfeger, M. and Bach, R.~L. (2024).
\newblock How does broadband supply affect participation in panel surveys?: {{Using}} geospatial broadband data at the district level to analyze mode choice and panel attrition.
\newblock {\em Quality \& Quantity}, 58(6):5805--5828.

\bibitem[{StataCorp}, 2023]{statacorpStataStatisticalSoftware2023}
{StataCorp} (2023).
\newblock Stata {{Statistical Software}}: {{Release}} 18.
\newblock College Station, TX: StataCorp LLC.

\bibitem[Struminskaya, 2014]{struminskayaDataQualityProbabilitybased2014}
Struminskaya, B. (2014).
\newblock {\em Data Quality in Probability-Based Online Panels: {{Nonresponse}}, Attrition, and Panel Conditioning}.
\newblock PhD thesis, Utrecht University.

\bibitem[Sturgis et~al., 2021]{sturgisInterviewerContributionVariability2021}
Sturgis, P., Maslovskaya, O., Durrant, G., and {Brunton-Smith}, I. (2021).
\newblock The {{Interviewer Contribution}} to {{Variability}} in {{Response Times}} in {{Face-to-Face Interview Surveys}}.
\newblock {\em Journal of Survey Statistics and Methodology}, 9(4):701--721.

\bibitem[Tortora, 2009]{tortoraAttritionConsumerPanels2009}
Tortora, R.~D. (2009).
\newblock Attrition in {{Consumer Panels}}.
\newblock In {\em Methodology of {{Longitudinal Surveys}}}, pages 235--249.

\bibitem[Turner et~al., 2015]{turnerCanResponseLatencies2015c}
Turner, G., Sturgis, P., and Martin, D. (2015).
\newblock Can {{Response Latencies Be Used}} to {{Detect Survey Satisficing}} on {{Cognitively Demanding Questions}}?
\newblock {\em Journal of Survey Statistics and Methodology}, 3(1):89--108.

\bibitem[{van Buuren} and {Groothuis-Oudshoorn}, 2011]{vanbuurenMiceMultivariateImputation2011}
{van Buuren}, S. and {Groothuis-Oudshoorn}, K. (2011).
\newblock Mice: {{Multivariate Imputation}} by {{Chained Equations}} in {{R}}.
\newblock {\em Journal of Statistical Software}, 45(3):1--67.

\bibitem[Williams, 2022a]{williamsHowWeBuilt2022}
Williams, J. (2022a).
\newblock How we built {{Public Voice}}: {{Kantar Public}}'s random sample panel.
\newblock {\em Social Research Practice}, (12):4--12.

\bibitem[Williams, 2022b]{williamsKantarPublicVoice2022}
Williams, J. (2022b).
\newblock Kantar {{Public Voice Panel Recruitment Surveys}} 1, 2 \& 3: {{Technical Report}} ({{Internal}}).
\newblock Technical report, Kantar Public.

\bibitem[Yan and Williams, 2022]{yanResponseBurdenReview2022}
Yan, T. and Williams, D. (2022).
\newblock Response {{Burden}} -- {{Review}} and {{Conceptual Framework}}.
\newblock {\em Journal of Official Statistics}, 38(4):939--961.

\end{thebibliography}
\end{document}